\documentclass[prb,twocolumn,floatfix,showpacs,amsmath,amssymb]{revtex4}
\usepackage{bbm}
\usepackage{amsfonts}
\usepackage{graphicx}
\usepackage{color}
\usepackage{amsmath}
\usepackage{amssymb}
\usepackage{latexsym}
\usepackage{psfrag}

\begin{document}

\preprint{Xu et al.}

\title{Electronic properties of quantum dots formed by magnetic double barriers in quantum wires}

\author{Hengyi Xu}\author{T. Heinzel}
\email{thomas.heinzel@uni-duesseldorf.de}
\affiliation{Solid State Physics Laboratory, Heinrich-Heine-Universit\"at, 40204 D\"usseldorf, Germany}
\author{I. V. Zozoulenko}
\affiliation{Solid State Electronics, Department of Science and Technology, Link\"{o}ping
University, 60174 Norrk\"{o}ping, Sweden}
\date{\today}

\begin{abstract}
The transport through a quantum wire exposed to two magnetic spikes in series is modeled. We demonstrate that quantum dots can be formed this way which couple to the leads via magnetic barriers. Conceptually, all quantum dot states are accessible by transport experiments. The simulations show Breit-Wigner resonances in the closed regime, while Fano resonances appear as soon as one open transmission channel is present. The system allows to tune the dot's confinement potential from sub-parabolic to superparabolic by experimentally accessible parameters.
\end{abstract}


\maketitle

\section{Introduction}

Quantum dots (QDs) are quasi zero-dimensional semiconducting systems in which the de Broglie wavelength of the electrons at the Fermi level is comparable to the spatial extension of the confinement. \cite{Heiss2005} Research on quantum dots has evolved into a mature field over the past 20 years. Within the top-down approach, QDs are typically defined in a two-dimensional electron gas (2DEG) residing in a semiconductor heterostructure and can be tuned electrostatically by nanopatterned gate electrodes. \cite{Kouwenhoven1997} QD formation by magnetic confinement is a potential alternative showing some fascinating and quite different phenomena. \cite{LeeS2004} For example, Sim et al. have calculated the energy spectrum of a QD formed by circular symmetric magnetic steps oriented perpendicular to a 2DEG and identified the corresponding, rosette shaped classical trajectories, revealing that the character of the bound states is completely different from those found in electrostatically defined QDs. \cite{Sim1998} It has also been predicted that chaotic QDs with tunable Lyapunov exponents can be formed by suitable magnetic field patterns. \cite{Voros2003} However, it has remained unclear how these systems can be implemented and probed experimentally. One suggestion from the theory side is to embed the magnetic dot in a quantum wire and to extract the information on the dot's energy spectrum via conductance resonances. \cite{Sim2001,Reijniers2001b} Experimentally, magnetic dots have been formed by using the fringe field of a ferromagnet on top of a semiconductor heterostructure. \cite{Nogaret2010} Magnetically bound, zero-dimensional states localized at the circumference of a magnetic coral have been observed on such samples. These states cause magnetoresistance oscillations which reflect the fact that scattering at the magnetic step is perfectly specular. \cite{Uzur2004} This magnetic confinement concept, however, leads to open dots which are strongly coupled to the environment such that Coulomb blockade is absent. This implies energy levels of large width and a poorly defined electron number in the dot. Moreover, the structures are large enough to allow an interpretation of the results within semiclassical pictures. We are not aware of a scheme for a purely magnetically confined QD which is weakly coupled to the environment. Strong confinement can, however, be achieved by combining electrostatic with magnetic fields. The basic idea is to confine electrons to a quantum wire with electrostatic walls and form a QD by adding a suitable magnetic field profile in the longitudinal direction. This approach has been studied theoretically by Reijniers et al. \cite{Reijniers2002} in the form of a quantum wire exposed to a longitudinal magnetic step where quantum states without a classical analogue have been predicted. This concept has been realized experimentally recently by a single magnetic barrier on top of a quantum wire and a superimposed homogeneous magnetic field. \cite{Tarasov2010} However, the observed bound states show strong coupling to the environment, while the weakly coupled states at lower energies are conceptually inaccessible by transport experiments due to the diamagnetic shifts in the leads.

\begin{figure}[tbp]
\includegraphics[width=85mm]{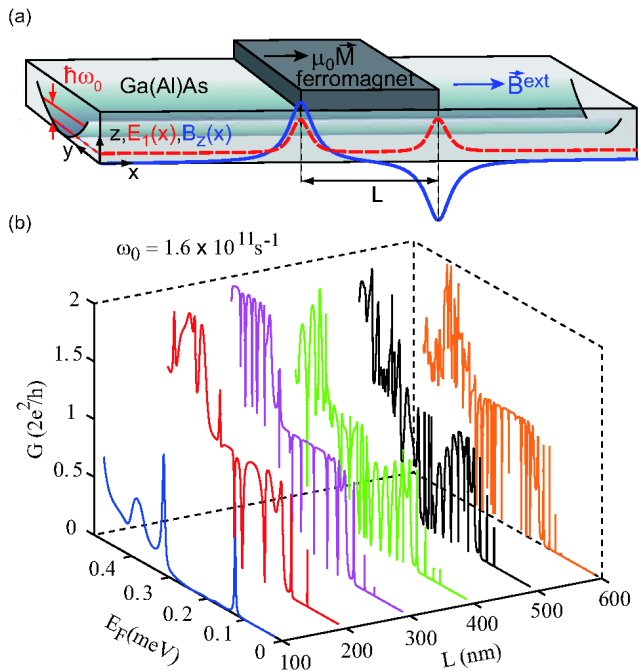}
\caption{(color online) (a) Scheme of the system under study. The parabolic quantum wire resides in a semiconductor, e.g. a $\mathrm{GaAs/Al_xGa_{1-x}As}$ heterostructure. A ferromagnetic stripe on top, magnetized in \textit{x}- direction by an external magnetic field $\mathbf{B}^{ext}$, generates an inhomogeneous, perpendicular magnetic fringe field $B_z(x)$ (blue solid line) composed of two magnetic barriers below the edges of the ferromagnet. The red, dashed line indicates the corresponding energy of the wire mode with the lowest energy, $E_1(x)$. (b) Conductance of the magnetoelectric quantum dot as a function of the Fermi energy $E_F$ and of the dot length $L$.} \label{DMBTheoryFig1}
\end{figure}

In the present paper, we  study theoretically QDs formed in a quantum wire which is exposed to two magnetic spikes (refereed to as \emph{magnetic barriers}) in series. It is demonstrated that in this system, strong magnetic confinement can be achieved where all states are experimentally accessible via transport measurements. Resonant tunneling dominates the transmission spectrum in the closed regime, while Fano resonances are found in the open regime. Furthermore, by changing the amplitude or the spacing of the magnetic barriers, or the distance of the 2DEG from the ferromagnetic stripe, the shape of the confinement potential can be tuned.

Several authors have previously calculated the conductance of two magnetic barriers in series which confine electrons in a quantum wire, and resonances indicative of the energy spectrum of the resulting island were predicted. \cite{Governale2000,Xu2001,Zhai2002,Guo2002} These structures have also been suggested as spin filters. \cite{Lu2002a,Seo2004,Jalil2005,Lu2005,Zhai2006,Scheid2007} However, in none of these papers, the properties of the QD itself or the character of the transmission resonances is studied, while a comprehensive semiclassical theory of this structure has been published recently.\cite{Papp2010}

The outline of the paper is a s follows. In section II, the model system and our numerical method are introduced. The results are presented and discussed in Section III. The paper closes with a summary and some conclusions (Section IV).

\section{Description of the model}

Our model system is sketched in Fig. \ref{DMBTheoryFig1} (a). It consists of a parabolic quantum wire (QWR) oriented in \textit{x}-direction with a parabolic confinement potential in \textit{y}-direction
\begin{equation}
V(y) = \frac{1}{2}m^{*}\omega_0^2y^2
\end{equation}
where $\omega_0=1.6\times 10^{11}s^{-1}$ is the confinement strength and $m^*=0.067 m_e$ the effective electron mass in GaAs. The QWR width $w$ in \textit{y}-direction depends on the Fermi energy $E_F$ which can be tuned, i.e. $w /E_F)=2\sqrt{E_F/m^*}/\omega_0$. The magnetic barriers are assumed to originate from a ferromagnetic stripe oriented across the QWR, i.e., in y-direction, which is magnetized in x direction by, e.g., a homogeneous, longitudinal external magnetic field. The perpendicular (\textit{z}-) component of the fringe field $B_z(x)$ forms two magnetic barriers in series of opposite polarity and with a spacing $L$ given by the width of the ferromagnetic stripe.\cite{Kubrak1999} This magnetic field profile can be written as \cite{Vancura2000}
\begin{widetext}
\begin{equation}
B_z(x) = \frac{\mu_0 M}{4\pi}\left[\ln\frac{(x+L/2)^2+d^2}{(x+L/2)^2+(d+h)^2}-\ln\frac{(x-L/2)^2+d^2}{(x-L/2)^2+(d+h)^2}\right]
\end{equation}
\end{widetext}
where $h$ denotes the thickness of ferromagnetic film and $d$ the distance of the QWR to the semiconductor surface. This magnetic field profile is depicted as well. The main effect on the QWR is provided by $B_z(x)$, while the in-plane components of $\mathbf{B}$ generate additional, small diamagnetic shifts which we neglect here. The effective g factor is set to zero and a spin degeneracy of 2 is assumed for all states. The parameters are geared to experimentally achievable values. One way to implement this geometry could be by placing a ferromagnetic stripe of submicron width \cite{Kubrak1999} across a quantum wire in a $\mathrm{GaAs/Al_xGa_{1-x}As}$ heterostructure. \cite{Hugger2008,Tarasov2010} The ferromagnetic film has an in-plane magnetization $\mu_0M$ in \textit{x}-direction, which we set to $2\,\mathrm{T}$. Magnetic barrier peak fields of $B_z^{max}\approx 0.57\mathrm{T}$ have been achieved experimentally this way. \cite{Hugger2007} Furthermore, we assume $h=60\,\mathrm{nm}$ and $d=30\,\mathrm{nm}$. An additional, homogeneous and nonmagnetic top gate may be used to tune the Fermi energy. \\

Qualitatively, the QWR modes experience an \textit{x}-dependent diamagnetic shift, and we denote their energies by $E_j(x)$ (j=1,2,...), given by \cite{Datta1997}
\begin{equation}
E_j(x)=(j-\frac{1}{2})\hbar \sqrt{\omega_0^2+\omega_c^2(x)}
\end{equation}
with the local cyclotron frequency $\omega_c(x)=eB_z(x)/m^{*}$.
A symmetric double barrier emerges as sketched for the first mode in Fig. \ref{DMBTheoryFig1}.

Quantitatively, the system is described by the effective-mass Hamiltonian
\begin{equation}
H = H_0 + V(y)
\end{equation}

where $H_0$ is the
kinetic energy term and $V(y)$ denotes the parabolic confinement potential defined in Eq. 1. The magnetic field $B_z(x)$ enters
via the vector potential as $\mathbf{A}=(-B_z(x)y, 0,0)$ and the Peierls substitution in the momentum operator $\mathbf{p}\rightarrow\mathbf{p}+e\mathbf{A}$.

The resulting Schr\"odinger equation is solved numerically on a discretized lattice with a lattice constants $a =2\,\mathrm{nm}$ in \textit{x}- direction. The discrete spatial coordinates at a given energy are thus $x=ma$ ($m=-500,-499,...500$). The tight-binding Hamiltonian of the
system reads
\begin{widetext}
\begin{eqnarray}
H = \sum_m \left\{\sum_n{\epsilon_0}c^\dag_{m,n}c_{m,n}-t\{c^%
\dag_{m,n}c_{m,n+1} +e^{-iqw}c^\dag_{m,n}c_{m+1,n}+\mathrm{H.c.}\}
\right\}
\end{eqnarray}
\end{widetext}
where $\epsilon_0$ is the site energy, $t=\hbar^2/(2m^*a^2)$ is the hopping matrix element and $c^\dag_{m,n} (c_{m,n})$ denotes the creation (annihilation) operators at site $(m,n)$, respectively, where $n$ is the index of the site in \textit{y}- direction, i.e. $y=nb$ with $b=2\,\mathrm{nm}$ and $n=0$ for $y=0$. Furthermore, the phase factor is given by $q=\frac{e}{\hbar}\int ^{x_{i+1}}_{x_i}B_z(x^{\prime}) dx^{\prime}$.
For transport calculations, the ends of the wire are connected to
ideal semi-infinite leads.
The two-terminal conductance $G$  is calculated within the Landauer-B\"{u}ttiker formalism  \cite{Datta1997}
\begin{equation}
G=\frac{2e^{2}}{h}\sum_{\beta ,\alpha =1}^{N}|t_{\beta \alpha
}|^{2}
\end{equation}
where $N$ is the number of propagating states in the leads,
$t_{\beta \alpha }$ is the transmission amplitude from incoming
state $\alpha $ in the left lead (at $x<-500a$) to outgoing state $\beta$ at $x>500a$. It can be expressed in terms of the
total Green's function $\mathcal{G}$ of
the system as $t_{\beta \alpha }=i\hbar \sqrt{v_{\alpha }v_{\beta }}\mathcal{G}^{501,-501}$, where $\mathcal{G}^{501,-501}$ denotes the matrix $\langle 501|%
\mathbf{\Gamma}|-501\rangle $ with $\pm 501$ corresponding to the
position of the left and right lead, respectively. We calculate
$\mathcal{G}$ using the recursive Green's function
technique in the hybrid space formulation.
\cite{Zozoulenko1996a,Zozoulenko1996b} Afterwards, we determine the
surface Green's functions related to the left and right leads and
the Green's function of the QWR separately and then link them at their interfaces.

The local density of states (LDOS) as a function of the site $\mathbf{r}=(m,n)$ is
related to the total Green's function in real space representation
by \cite{Datta1997}
\begin{equation}
LDOS (\mathbf{r};E)=-\frac{1}{\pi
}Im[\mathbf{\Gamma}(\mathbf{r,r};E)]
\end{equation}%
where $\emph{Im}$ denotes the imaginary part.

\section{Results and Discussion}

\begin{figure}[tbp]
\includegraphics[width=85mm]{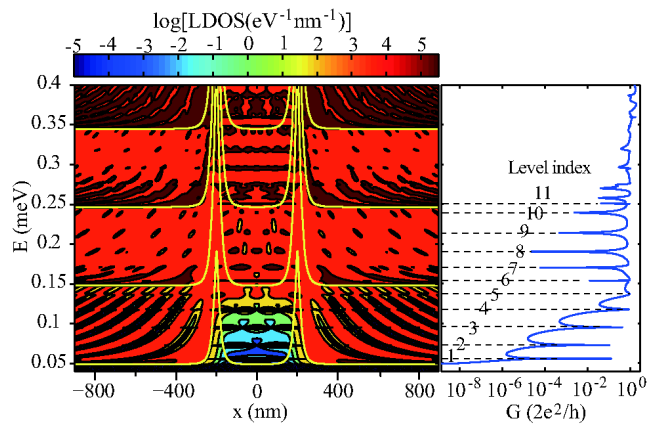}
\caption{(color online) Left: The LDOS as a function of $E_F$ and x. Also shown (yellow bold lines) are the mode energies $E_j(x)$. The magnetic barriers reside at $x=\pm200\,\mathrm{nm}$. The corresponding two-terminal conductance is shown to the right. } \label{DMBTheoryFig2}
\end{figure}

Fig. \ref{DMBTheoryFig1} (b) shows an overview of the conductance through the QD as a function of $L$ and $E_F$. Most prominently, transmission dips are seen in the open regime ($G\geq 2e^2/h$), superimposed to the quantized conductance plateaux. The spectral density of the resonances increases with increasing $L$ and (not shown) as $\omega_0$ is decreased. This indicates that the dips are related to the energy spectrum of the dot and not to effects at the individual magnetic barriers as reported in Ref. \onlinecite{Xu2007a} (note that the results reported there have been obtained for hard wall wires and get suppressed in softer confinement potentials). In the closed regime, i.e., for $G<2e^2/h$, transmission resonances can be identified.

\begin{figure}[tbp]
\includegraphics[width=85mm]{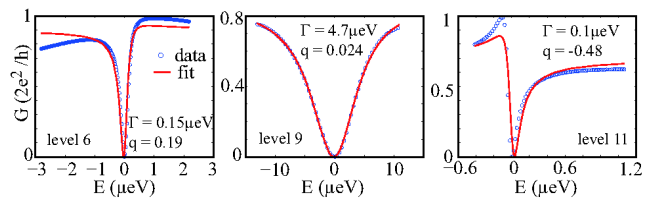}
\caption{Fits of selected resonances shown in Fig. \ref{DMBTheoryFig2} to the Fano formula.} \label{DMBTheoryFig3}
\end{figure}

We now focus on a dot of length $L=400\,\mathrm{nm}$ in the low-energy regime where at most the lowest three modes are occupied in the leads. In Fig. \ref{DMBTheoryFig2}, the LDOS as a function of $x$ and $E_F$, integrated along the $y$-direction, is shown for a magnetization of $\mu_0M=2\,\mathrm{T}$. As expected from the qualitative picture sketched above, the diamagnetic shifts of the mode energies form an effective double barrier (yellow bold lines) leading to quantized confinement. In the single mode regime ($50\,\mathrm{\mu eV}<E_F<150\,\mathrm{\mu eV}$), the mode spacing increases with increasing energy, reflecting the superparabolic shape of the confinement potential in longitudinal direction. The conductance correlates with the dot spectrum and shows resonant tunneling peaks in this regime (right part of Fig. \ref{DMBTheoryFig2}). As the second QWR mode gets occupied, the character of the resonances switches from transmissive to reflective. In contrast to the transmission resonances in the closed regime, these resonances are not necessarily symmetric. They originate from interferences of the propagating states of the first QWR modes with bound states of the second or third QWR mode and can thus be described by the Fano line shape \cite{Fano1961}
\begin{equation}
G(E_F)=\frac{2e^2}{h}\frac{1}{1+q^2}\cdot\frac{[q\pm\epsilon/2\Gamma]^2}{1+[\epsilon/2\Gamma]^2}
\label{EqnFanoLineShape}
\end{equation}

Here, $\Gamma$ denotes the coupling of the bound state to the leads, $\epsilon=E_F-E_{res}$ is the detuning from the resonance center $E_{res}$, and $q$ is the Fano parameter which is a measure of the phase difference between the two transmission channels the electron waves collect as they traverse the dot, i.e., $q=-\cot{[(\alpha-\delta)/2]}$, where $\alpha$ ($\delta$) denotes the phase the wave acquires while traversing the dot via the bound (free) state. For $q=0$, a symmetric dip is obtained. As $|q|$ increases, the line shape gets asymmetric, and the limit $|q|\rightarrow \infty$ reduces Eq. \ref{EqnFanoLineShape} to a Breit-Wigner resonance. Furthermore, for $q>0$ ($q<0)$, the dip appears to the left (right) of a peak.\\

Fig. \ref{DMBTheoryFig3} depicts the fits of some resonances to Eq. \eqref{EqnFanoLineShape}.  We find both positive and negative Fano parameters in the range $|q|<0.5$, while the coupling of the states to the leads varies by up to a factor of 50. This behavior resembles that one observed by Gores et al. \cite{Gores2000} on an open, electrostatically defined QD, where, however, the Fano resonances showed larger asymmetries and a more homogeneous coupling. The deviations between fit and simulated data can be attributed to distortions of nearby resonances.

\begin{figure}[tbp]
\includegraphics[width=85mm]{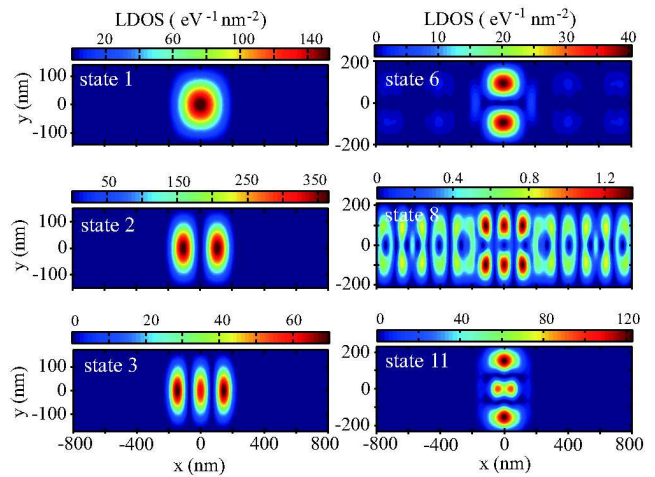}
\caption{(color online) The LDOS of the bound states at the energies of selected resonances. The labels of the states refer to Fig. 2.} \label{DMBTheoryFig4}
\end{figure}

The character of the bound states and their coupling to the leads can be visualized with the help of spatially resolved LDOS plots, see Fig. \ref{DMBTheoryFig4}, where the LDOS as a function of x and y is plotted for selected energy levels. States belonging to the first wire mode have no node in y-direction, and the level index is given by the number of nodes in x-direction. State 6 (11) is the lowest energy state belonging to the second (third) wire mode. While sharp resonances originate from states well localized close to the center of the dot, transmission via more extended states, e.g. state 8, causes broad and highly symmetric resonances.

\begin{figure}[tbp]
\includegraphics[width=75mm]{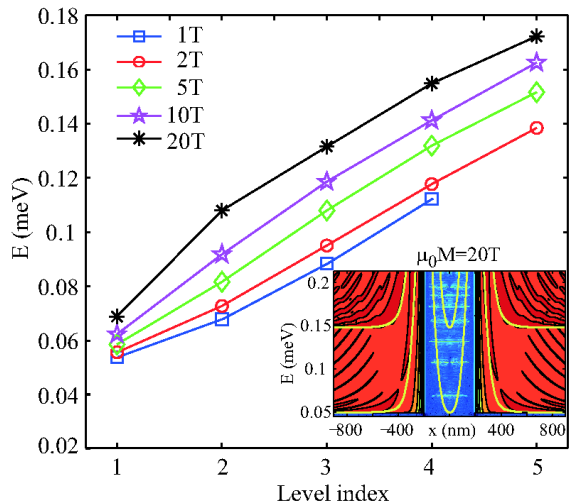}
\caption{(color online) Parametric tuning of the confinement potential shape. Main figure: energies of the quantum dot states in the closed regime as a function of the level index for various MB amplitudes. Inset: LDOS ($E_F, x$) and $E_1(x)$ for a MB amplitude of $20\,\mathrm{T}$, showing a sub-parabolic confinement effective potential. } \label{DMBTheoryFig5}
\end{figure}

One feature with potential experimental relevance of this system is that it allows to tune the confinement potential shape by experimentally controllable parameters, i.e., by the spacing $L$ or the amplitude $B_z^{max}$ of the magnetic barrier, or the distance $d$ of the electron gas from the surface. In general, the less the two magnetic barriers overlap, the steeper the effective confinement potential gets. While changing $L$ requires fabrication of several samples and $d$ can only be tuned by about $25\,\mathrm{nm}$ in, e.g., parabolic quantum wells embedded between a top and a back gate,\cite{Salis1997} $B_z^{max}$ can be tuned over wide ranges by changing $\mu_0M$. Therefore, we exemplify this tunability by using $\mu_0M$ as parameter. In Fig. \ref{DMBTheoryFig5}, the energies of the bound states in the closed regime are plotted as a function of the level index $\ell$ for various magnetic fields. For barrier amplitudes $\mu_0M<5\,\mathrm{T}$, the level spacing increases with $\ell$ indicating superparabolic confinement, while for $\mu_0M>5\,\mathrm{T}$, the confinement is sub-parabolic. For the sample parameters chosen here, a magnetization of $5\,\mathrm{T}$ represents an approximately parabolic confinement. Note that the case $B_z^{max}=2\,\mathrm{T}$ is plotted in Fig. 2. The LDOS as a function of $E_F$ and $x$ for the strongest magnetization is shown in the inset of Fig. \ref{DMBTheoryFig5}, where the sub-parabolicity is directly visible.

\section{Summary and Conclusions}
We have shown that by exposing a quantum wire to two magnetic barriers in series, tunable quantum dots can be formed in which all discrete states are conceptually accessible by transport experiments. Such quantum dots may provide an alternative to conventional quantum dots defined by purely electrostatic confinement. They reveal the full variety of the possible forms of Fano resonances and show a large variation of the coupling of the dot states to the leads. Furthermore, the shape of the confinement potential can be changed continuously between subparabolic and superparabolic by experimentally accessible parameters.

Quantum effects at magnetic double barriers in quantum wires have not been tested experimentally yet to the best of our knowledge. This is somewhat surprising, in particular considering that the necessary technology is essentially established. Several experiments on magnetic double barriers with sub-micron spacing have been performed in wide 2DEGs,  \cite{Kubrak1999,Joo2006,Bae2007,Lin2010} with the minimum 2DEG width of $1\mu m$, reported in Ref. \onlinecite{Joo2006}, which is probably too broad for detecting quantized states. Suitable magnetic field profiles have been generated by using nonplanar 2DEGs \cite{Leadbeater1995} as well as superconductor- \cite{Geim1992,Carmona1995} or ferromagnet- \cite{Ye1995,Johnson1997,Monzon1997,Kubrak1999,Vancura2000,Cerchez2007} semiconductor hybrid structures. We note that the experimentally available magnetizations are limited to about $3.75\,\mathrm{T}$, such that the sub-parabolic potential regime discussed above lies outside experimental reach, but it should be possible to scale the system accordingly by a careful choice of the sample parameters. Furthermore, stronger magnetic barriers can be achieved by the sample layout used in Ref. \onlinecite{Leadbeater1995}. As  an outlook, we mention that magnetic confinement concepts have attained increased attention recently due to their potential application to graphene, \cite{Martino2007} where electrostatic confinement is inhibited due to Klein tunneling. \cite{Castro2009} It will be interesting to see whether the concept discussed here can be transferred to graphene nanoribbons.

\begin{acknowledgments}
H.X. and T.H. acknowledge financial support from Heinrich-Heine-Universit\"{a}t D\"{u}sseldorf.
\end{acknowledgments}


\begin{thebibliography}{45}
\expandafter\ifx\csname natexlab\endcsname\relax\def\natexlab#1{#1}\fi
\expandafter\ifx\csname bibnamefont\endcsname\relax
  \def\bibnamefont#1{#1}\fi
\expandafter\ifx\csname bibfnamefont\endcsname\relax
  \def\bibfnamefont#1{#1}\fi
\expandafter\ifx\csname citenamefont\endcsname\relax
  \def\citenamefont#1{#1}\fi
\expandafter\ifx\csname url\endcsname\relax
  \def\url#1{\texttt{#1}}\fi
\expandafter\ifx\csname urlprefix\endcsname\relax\def\urlprefix{URL }\fi
\providecommand{\bibinfo}[2]{#2}
\providecommand{\eprint}[2][]{\url{#2}}

\bibitem[{\citenamefont{(ed.)}(2005)}]{Heiss2005}
\bibinfo{author}{\bibfnamefont{D.~H.} \bibnamefont{(ed.)}},
  \emph{\bibinfo{title}{\emph{Quantum dots: a doorway to nanoscale physics}}}
  (\bibinfo{publisher}{Springer, Berlin}, \bibinfo{year}{2005}).

\bibitem[{\citenamefont{Kouwenhoven et~al.}(1997)\citenamefont{Kouwenhoven,
  Marcus, McEuen, Tarucha, Westervelt, and Wingreen}}]{Kouwenhoven1997}
\bibinfo{author}{\bibfnamefont{L.~P.} \bibnamefont{Kouwenhoven}},
  \bibinfo{author}{\bibfnamefont{C.~M.} \bibnamefont{Marcus}},
  \bibinfo{author}{\bibfnamefont{P.~L.} \bibnamefont{McEuen}},
  \bibinfo{author}{\bibfnamefont{S.}~\bibnamefont{Tarucha}},
  \bibinfo{author}{\bibfnamefont{R.~M.} \bibnamefont{Westervelt}},
  \bibnamefont{and} \bibinfo{author}{\bibfnamefont{N.~S.}
  \bibnamefont{Wingreen}}, in \emph{\bibinfo{booktitle}{Mesoscopic Electron
  Transport, Series E: Applied Sciences}}, edited by
  \bibinfo{editor}{\bibfnamefont{L.~L.} \bibnamefont{Sohn}},
  \bibinfo{editor}{\bibfnamefont{L.~P.} \bibnamefont{Kouwenhoven}},
  \bibnamefont{and}
  \bibinfo{editor}{\bibfnamefont{G.}~\bibnamefont{Sch$\rm{\ddot{o}}$n}}
  (\bibinfo{publisher}{Kluwer}, \bibinfo{year}{1997}).

\bibitem[{\citenamefont{Lee}(2004)}]{LeeS2004}
\bibinfo{author}{\bibfnamefont{S.~J.~.} \bibnamefont{Lee}},
  \bibinfo{journal}{Phys. Rep.} \textbf{\bibinfo{volume}{394}},
  \bibinfo{pages}{1} (\bibinfo{year}{2004}).

\bibitem[{\citenamefont{Sim et~al.}(1998)\citenamefont{Sim, Ahn, Chang, Ihm,
  Kim, and Lee}}]{Sim1998}
\bibinfo{author}{\bibfnamefont{H.-S.} \bibnamefont{Sim}},
  \bibinfo{author}{\bibfnamefont{K.~H.} \bibnamefont{Ahn}},
  \bibinfo{author}{\bibfnamefont{K.~J.} \bibnamefont{Chang}},
  \bibinfo{author}{\bibfnamefont{G.}~\bibnamefont{Ihm}},
  \bibinfo{author}{\bibfnamefont{N.}~\bibnamefont{Kim}}, \bibnamefont{and}
  \bibinfo{author}{\bibfnamefont{S.~J.} \bibnamefont{Lee}},
  \bibinfo{journal}{Phys. Rev. Lett.} \textbf{\bibinfo{volume}{80}},
  \bibinfo{pages}{1501} (\bibinfo{year}{1998}).

\bibitem[{\citenamefont{V$\rm{\ddot{o}}$r$\rm{\ddot{o}}$s
  et~al.}(2003)\citenamefont{V$\rm{\ddot{o}}$r$\rm{\ddot{o}}$s, Tasnadi,
  Cserti, and Pollner}}]{Voros2003}
\bibinfo{author}{\bibfnamefont{Z.}~\bibnamefont{V$\rm{\ddot{o}}$r$\rm{\ddot{o}%
}$s}}, \bibinfo{author}{\bibfnamefont{T.}~\bibnamefont{Tasnadi}},
  \bibinfo{author}{\bibfnamefont{J.}~\bibnamefont{Cserti}}, \bibnamefont{and}
  \bibinfo{author}{\bibfnamefont{P.}~\bibnamefont{Pollner}},
  \bibinfo{journal}{Phys. Rev. E} \textbf{\bibinfo{volume}{67}},
  \bibinfo{pages}{065202} (\bibinfo{year}{2003}).

\bibitem[{\citenamefont{Sim et~al.}(2001)\citenamefont{Sim, Ihm, Kim, and
  Chang}}]{Sim2001}
\bibinfo{author}{\bibfnamefont{H.-S.} \bibnamefont{Sim}},
  \bibinfo{author}{\bibfnamefont{G.}~\bibnamefont{Ihm}},
  \bibinfo{author}{\bibfnamefont{N.}~\bibnamefont{Kim}}, \bibnamefont{and}
  \bibinfo{author}{\bibfnamefont{K.~J.} \bibnamefont{Chang}},
  \bibinfo{journal}{Phys. Rev. Lett.} \textbf{\bibinfo{volume}{87}},
  \bibinfo{pages}{146601} (\bibinfo{year}{2001}).

\bibitem[{\citenamefont{Reijniers and Matulis}(2001)}]{Reijniers2001b}
\bibinfo{author}{\bibfnamefont{J.}~\bibnamefont{Reijniers}},
  \bibinfo{author}{\bibfnamefont{F.~M.}~\bibnamefont{Peeters}}, \bibnamefont{and} \bibinfo{author}{\bibfnamefont{A.} \bibnamefont{Matulis}},
  \bibinfo{journal}{Phys. Rev. B} \textbf{\bibinfo{volume}{64}},
  \bibinfo{pages}{245314} (\bibinfo{year}{2001}).

\bibitem[{\citenamefont{Nogaret}(2010)}]{Nogaret2010}
\bibinfo{author}{\bibfnamefont{A.}~\bibnamefont{Nogaret}}, \bibinfo{journal}{J.
  Phys. Cond. Mat.} \textbf{\bibinfo{volume}{22}}, \bibinfo{pages}{253201}
  (\bibinfo{year}{2010}).

\bibitem[{\citenamefont{Uzur et~al.}(2004)\citenamefont{Uzur, Nogaret, Beere,
  Ritchie, Marrows, and Hickey}}]{Uzur2004}
\bibinfo{author}{\bibfnamefont{D.}~\bibnamefont{Uzur}},
  \bibinfo{author}{\bibfnamefont{A.}~\bibnamefont{Nogaret}},
  \bibinfo{author}{\bibfnamefont{H.~E.} \bibnamefont{Beere}},
  \bibinfo{author}{\bibfnamefont{D.~A.} \bibnamefont{Ritchie}},
  \bibinfo{author}{\bibfnamefont{C.~H.} \bibnamefont{Marrows}},
  \bibnamefont{and} \bibinfo{author}{\bibfnamefont{B.~J.}
  \bibnamefont{Hickey}}, \bibinfo{journal}{Phys. Rev. B}
  \textbf{\bibinfo{volume}{69}}, \bibinfo{pages}{241301}
  (\bibinfo{year}{2004}).

\bibitem[{\citenamefont{Reijniers et~al.}(2002)\citenamefont{Reijniers,
  Matulis, Chang, Peeters, and Vasilopoulos}}]{Reijniers2002}
\bibinfo{author}{\bibfnamefont{J.}~\bibnamefont{Reijniers}},
  \bibinfo{author}{\bibfnamefont{A.}~\bibnamefont{Matulis}},
  \bibinfo{author}{\bibfnamefont{K.}~\bibnamefont{Chang}},
  \bibinfo{author}{\bibfnamefont{F.~M.} \bibnamefont{Peeters}},
  \bibnamefont{and}
  \bibinfo{author}{\bibfnamefont{P.}~\bibnamefont{Vasilopoulos}},
  \bibinfo{journal}{Europhys. Lett.} \textbf{\bibinfo{volume}{59}},
  \bibinfo{pages}{749} (\bibinfo{year}{2002}).

\bibitem[{\citenamefont{Tarasov et~al.}(2010)\citenamefont{Tarasov, Hugger, Xu,
  Cerchez, Heinzel, Zozoulenko, Gasser-Szerer, Reuter, and
  Wieck}}]{Tarasov2010}
\bibinfo{author}{\bibfnamefont{A.}~\bibnamefont{Tarasov}},
  \bibinfo{author}{\bibfnamefont{S.}~\bibnamefont{Hugger}},
  \bibinfo{author}{\bibfnamefont{H.}~\bibnamefont{Xu}},
  \bibinfo{author}{\bibfnamefont{M.}~\bibnamefont{Cerchez}},
  \bibinfo{author}{\bibfnamefont{T.}~\bibnamefont{Heinzel}},
  \bibinfo{author}{\bibfnamefont{I.~V.} \bibnamefont{Zozoulenko}},
  \bibinfo{author}{\bibfnamefont{U.}~\bibnamefont{Gasser-Szerer}},
  \bibinfo{author}{\bibfnamefont{D.}~\bibnamefont{Reuter}}, \bibnamefont{and}
  \bibinfo{author}{\bibfnamefont{A.~D.} \bibnamefont{Wieck}},
  \bibinfo{journal}{Phys. Rev. Lett.} \textbf{\bibinfo{volume}{104}},
  \bibinfo{pages}{186801} (\bibinfo{year}{2010}).

\bibitem[{\citenamefont{Governale and Boese}(2000)}]{Governale2000}
\bibinfo{author}{\bibfnamefont{M.}~\bibnamefont{Governale}} \bibnamefont{and}
  \bibinfo{author}{\bibfnamefont{D.}~\bibnamefont{Boese}},
  \bibinfo{journal}{Appl. Phys. Lett.} \textbf{\bibinfo{volume}{77}},
  \bibinfo{pages}{3215} (\bibinfo{year}{2000}).

\bibitem[{\citenamefont{Xu and Okada}(2001)}]{Xu2001}
\bibinfo{author}{\bibfnamefont{H.~Z.} \bibnamefont{Xu}} \bibnamefont{and}
  \bibinfo{author}{\bibfnamefont{Y.}~\bibnamefont{Okada}},
  \bibinfo{journal}{Appl. Phys. Lett.} \textbf{\bibinfo{volume}{79}},
  \bibinfo{pages}{3119} (\bibinfo{year}{2001}).

\bibitem[{\citenamefont{Zhai et~al.}(2002)\citenamefont{Zhai, Guo, and
  Gu}}]{Zhai2002}
\bibinfo{author}{\bibfnamefont{F.}~\bibnamefont{Zhai}},
  \bibinfo{author}{\bibfnamefont{Y.}~\bibnamefont{Guo}}, \bibnamefont{and}
  \bibinfo{author}{\bibfnamefont{B.-L.} \bibnamefont{Gu}},
  \bibinfo{journal}{Phys. Rev. B} \textbf{\bibinfo{volume}{66}},
  \bibinfo{pages}{125305} (\bibinfo{year}{2002}).

\bibitem[{\citenamefont{Guo et~al.}(2002)\citenamefont{Guo, Zhai, Gu, and
  Kawazoe}}]{Guo2002}
\bibinfo{author}{\bibfnamefont{Y.}~\bibnamefont{Guo}},
  \bibinfo{author}{\bibfnamefont{F.}~\bibnamefont{Zhai}},
  \bibinfo{author}{\bibfnamefont{B.~L.} \bibnamefont{Gu}}, \bibnamefont{and}
  \bibinfo{author}{\bibfnamefont{Y.}~\bibnamefont{Kawazoe}},
  \bibinfo{journal}{Phys. Rev. B} \textbf{\bibinfo{volume}{66}},
  \bibinfo{pages}{045312} (\bibinfo{year}{2002}).

\bibitem[{\citenamefont{Lu et~al.}(2002)\citenamefont{Lu, Zhang, and
  Yan}}]{Lu2002a}
\bibinfo{author}{\bibfnamefont{M.-W.} \bibnamefont{Lu}},
  \bibinfo{author}{\bibfnamefont{L.-D.} \bibnamefont{Zhang}}, \bibnamefont{and}
  \bibinfo{author}{\bibfnamefont{X.-H.} \bibnamefont{Yan}},
  \bibinfo{journal}{Phys. Rev. B} \textbf{\bibinfo{volume}{66}},
  \bibinfo{pages}{224412} (\bibinfo{year}{2002}).

\bibitem[{\citenamefont{Seo et~al.}(2004)\citenamefont{Seo, Ihm, Ahn, and
  Lee}}]{Seo2004}
\bibinfo{author}{\bibfnamefont{K.~C.} \bibnamefont{Seo}},
  \bibinfo{author}{\bibfnamefont{G.}~\bibnamefont{Ihm}},
  \bibinfo{author}{\bibfnamefont{K.-H.} \bibnamefont{Ahn}}, \bibnamefont{and}
  \bibinfo{author}{\bibfnamefont{S.~J.} \bibnamefont{Lee}},
  \bibinfo{journal}{J. Appl. Phys.} \textbf{\bibinfo{volume}{95}},
  \bibinfo{pages}{7252} (\bibinfo{year}{2004}).

\bibitem[{\citenamefont{Jalil}(2005)}]{Jalil2005}
\bibinfo{author}{\bibfnamefont{M.~B.~A.} \bibnamefont{Jalil}},
  \bibinfo{journal}{J. Appl. Phys.} \textbf{\bibinfo{volume}{97}},
  \bibinfo{pages}{024507} (\bibinfo{year}{2005}).

\bibitem[{\citenamefont{Lu}(2005)}]{Lu2005}
\bibinfo{author}{\bibfnamefont{J.-D.} \bibnamefont{Lu}},
  \bibinfo{journal}{Appl. Surf. Science} \textbf{\bibinfo{volume}{254}},
  \bibinfo{pages}{5044} (\bibinfo{year}{2005}).

\bibitem[{\citenamefont{Zhai and Xu}(2006)}]{Zhai2006}
\bibinfo{author}{\bibfnamefont{F.}~\bibnamefont{Zhai}} \bibnamefont{and}
  \bibinfo{author}{\bibfnamefont{H.~Q.} \bibnamefont{Xu}},
  \bibinfo{journal}{Appl. Phys. Lett.} \textbf{\bibinfo{volume}{88}},
  \bibinfo{pages}{032502} (\bibinfo{year}{2006}).

\bibitem[{\citenamefont{Scheid et~al.}(2007)\citenamefont{Scheid, Bercioux, and
  Richter}}]{Scheid2007}
\bibinfo{author}{\bibfnamefont{M.}~\bibnamefont{Scheid}},
  \bibinfo{author}{\bibfnamefont{D.}~\bibnamefont{Bercioux}}, \bibnamefont{and}
  \bibinfo{author}{\bibfnamefont{K.}~\bibnamefont{Richter}},
  \bibinfo{journal}{New J. of Physics} \textbf{\bibinfo{volume}{9}},
  \bibinfo{pages}{401} (\bibinfo{year}{2007}).

\bibitem[{\citenamefont{Papp and Peeters}(2010)}]{Papp2010}
\bibinfo{author}{\bibfnamefont{G.}~\bibnamefont{Papp}} \bibnamefont{and}
  \bibinfo{author}{\bibfnamefont{F.~M.} \bibnamefont{Peeters}},
  \bibinfo{journal}{J. Appl. Phys.} \textbf{\bibinfo{volume}{107}},
  \bibinfo{pages}{063718} (\bibinfo{year}{2010}).

\bibitem[{\citenamefont{Kubrak et~al.}(1999)\citenamefont{Kubrak, Rahman,
  Gallagher, Main, Henini, Marrows, and Howson}}]{Kubrak1999}
\bibinfo{author}{\bibfnamefont{V.}~\bibnamefont{Kubrak}},
  \bibinfo{author}{\bibfnamefont{F.}~\bibnamefont{Rahman}},
  \bibinfo{author}{\bibfnamefont{B.~L.} \bibnamefont{Gallagher}},
  \bibinfo{author}{\bibfnamefont{P.~C.} \bibnamefont{Main}},
  \bibinfo{author}{\bibfnamefont{M.}~\bibnamefont{Henini}},
  \bibinfo{author}{\bibfnamefont{C.~H.} \bibnamefont{Marrows}},
  \bibnamefont{and} \bibinfo{author}{\bibfnamefont{M.~A.}
  \bibnamefont{Howson}}, \bibinfo{journal}{Appl. Phys. Lett.}
  \textbf{\bibinfo{volume}{74}}, \bibinfo{pages}{2507} (\bibinfo{year}{1999}).

\bibitem[{\citenamefont{Van$\mathrm{\check{c}}$ura
  et~al.}(2000)\citenamefont{Van$\mathrm{\check{c}}$ura, Ihn, Broderick,
  Ensslin, Wegscheider, and Bichler}}]{Vancura2000}
\bibinfo{author}{\bibfnamefont{T.}~\bibnamefont{Van$\mathrm{\check{c}}$ura}},
  \bibinfo{author}{\bibfnamefont{T.}~\bibnamefont{Ihn}},
  \bibinfo{author}{\bibfnamefont{S.}~\bibnamefont{Broderick}},
  \bibinfo{author}{\bibfnamefont{K.}~\bibnamefont{Ensslin}},
  \bibinfo{author}{\bibfnamefont{W.}~\bibnamefont{Wegscheider}},
  \bibnamefont{and} \bibinfo{author}{\bibfnamefont{M.}~\bibnamefont{Bichler}},
  \bibinfo{journal}{Phys. Rev. B} \textbf{\bibinfo{volume}{62}},
  \bibinfo{pages}{5074} (\bibinfo{year}{2000}).

\bibitem[{\citenamefont{Hugger et~al.}(2008)\citenamefont{Hugger, Xu, Tarasov,
  Cerchez, Heinzel, Zozoulenko, Reuter, and Wieck}}]{Hugger2008}
\bibinfo{author}{\bibfnamefont{S.}~\bibnamefont{Hugger}},
  \bibinfo{author}{\bibfnamefont{H.}~\bibnamefont{Xu}},
  \bibinfo{author}{\bibfnamefont{A.}~\bibnamefont{Tarasov}},
  \bibinfo{author}{\bibfnamefont{M.}~\bibnamefont{Cerchez}},
  \bibinfo{author}{\bibfnamefont{T.}~\bibnamefont{Heinzel}},
  \bibinfo{author}{\bibfnamefont{I.~V.} \bibnamefont{Zozoulenko}},
  \bibinfo{author}{\bibfnamefont{D.}~\bibnamefont{Reuter}}, \bibnamefont{and}
  \bibinfo{author}{\bibfnamefont{A.~D.} \bibnamefont{Wieck}},
  \bibinfo{journal}{Phys. Rev. B} \textbf{\bibinfo{volume}{78}},
  \bibinfo{pages}{165307} (\bibinfo{year}{2008}).

\bibitem[{\citenamefont{Hugger et~al.}(2007)\citenamefont{Hugger, Cerchez, Xu,
  and Heinzel}}]{Hugger2007}
\bibinfo{author}{\bibfnamefont{S.}~\bibnamefont{Hugger}},
  \bibinfo{author}{\bibfnamefont{M.}~\bibnamefont{Cerchez}},
  \bibinfo{author}{\bibfnamefont{H.}~\bibnamefont{Xu}}, \bibnamefont{and}
  \bibinfo{author}{\bibfnamefont{T.}~\bibnamefont{Heinzel}},
  \bibinfo{journal}{Phys. Rev. B} \textbf{\bibinfo{volume}{76}},
  \bibinfo{pages}{195308} (\bibinfo{year}{2007}).

\bibitem[{\citenamefont{Datta}(1997)}]{Datta1997}
\bibinfo{author}{\bibfnamefont{S.}~\bibnamefont{Datta}},
  \emph{\bibinfo{title}{\emph{Electronic Transport in Mesoscopic Systems}}}
  (\bibinfo{publisher}{Cambridge University Press}, \bibinfo{year}{1997}).

\bibitem[{\citenamefont{Zozoulenko
  et~al.}(1996{\natexlab{a}})\citenamefont{Zozoulenko, Maao, and
  Hauge}}]{Zozoulenko1996a}
\bibinfo{author}{\bibfnamefont{I.~V.} \bibnamefont{Zozoulenko}},
  \bibinfo{author}{\bibfnamefont{F.~A.} \bibnamefont{Maao}}, \bibnamefont{and}
  \bibinfo{author}{\bibfnamefont{E.~H.} \bibnamefont{Hauge}},
  \bibinfo{journal}{Phys. Rev. B} \textbf{\bibinfo{volume}{53}},
  \bibinfo{pages}{7975} (\bibinfo{year}{1996}{\natexlab{a}}).

\bibitem[{\citenamefont{Zozoulenko
  et~al.}(1996{\natexlab{b}})\citenamefont{Zozoulenko, Maao, and
  Hauge}}]{Zozoulenko1996b}
\bibinfo{author}{\bibfnamefont{I.~V.} \bibnamefont{Zozoulenko}},
  \bibinfo{author}{\bibfnamefont{F.~A.} \bibnamefont{Maao}}, \bibnamefont{and}
  \bibinfo{author}{\bibfnamefont{E.~H.} \bibnamefont{Hauge}},
  \bibinfo{journal}{Phys. Rev. B} \textbf{\bibinfo{volume}{53}},
  \bibinfo{pages}{7987} (\bibinfo{year}{1996}{\natexlab{b}}).

\bibitem[{\citenamefont{Xu et~al.}(2007)\citenamefont{Xu, Heinzel, Evaldsson,
  Ihnatsenka, and Zozoulenko}}]{Xu2007a}
\bibinfo{author}{\bibfnamefont{H.}~\bibnamefont{Xu}},
  \bibinfo{author}{\bibfnamefont{T.}~\bibnamefont{Heinzel}},
  \bibinfo{author}{\bibfnamefont{M.}~\bibnamefont{Evaldsson}},
  \bibinfo{author}{\bibfnamefont{S.}~\bibnamefont{Ihnatsenka}},
  \bibnamefont{and} \bibinfo{author}{\bibfnamefont{I.~V.}
  \bibnamefont{Zozoulenko}}, \bibinfo{journal}{Phys. Rev. B}
  \textbf{\bibinfo{volume}{75}}, \bibinfo{pages}{205301}
  (\bibinfo{year}{2007}).

\bibitem[{\citenamefont{Fano}(1961)}]{Fano1961}
\bibinfo{author}{\bibfnamefont{U.}~\bibnamefont{Fano}}, \bibinfo{journal}{Phys.
  Rev.} \textbf{\bibinfo{volume}{124}}, \bibinfo{pages}{1866}
  (\bibinfo{year}{1961}).

\bibitem[{\citenamefont{Gores et~al.}(2000)\citenamefont{Gores,
  Goldhaber-Gordon, Heemeyer, Kastner, Shtrikman, Mahalu, and
  Meirav}}]{Gores2000}
\bibinfo{author}{\bibfnamefont{J.}~\bibnamefont{Gores}},
  \bibinfo{author}{\bibfnamefont{D.}~\bibnamefont{Goldhaber-Gordon}},
  \bibinfo{author}{\bibfnamefont{S.}~\bibnamefont{Heemeyer}},
  \bibinfo{author}{\bibfnamefont{M.~A.} \bibnamefont{Kastner}},
  \bibinfo{author}{\bibfnamefont{H.}~\bibnamefont{Shtrikman}},
  \bibinfo{author}{\bibfnamefont{D.}~\bibnamefont{Mahalu}}, \bibnamefont{and}
  \bibinfo{author}{\bibfnamefont{U.}~\bibnamefont{Meirav}},
  \bibinfo{journal}{Phys. Rev. B} \textbf{\bibinfo{volume}{62}},
  \bibinfo{pages}{2188} (\bibinfo{year}{2000}).

\bibitem[{\citenamefont{Salis et~al.}(1997)\citenamefont{Salis, Graf, Ensslin,
  Campman, Maranowski, and Gossard}}]{Salis1997}
\bibinfo{author}{\bibfnamefont{G.}~\bibnamefont{Salis}},
  \bibinfo{author}{\bibfnamefont{B.}~\bibnamefont{Graf}},
  \bibinfo{author}{\bibfnamefont{K.}~\bibnamefont{Ensslin}},
  \bibinfo{author}{\bibfnamefont{K.}~\bibnamefont{Campman}},
  \bibinfo{author}{\bibfnamefont{K.}~\bibnamefont{Maranowski}},
  \bibnamefont{and} \bibinfo{author}{\bibfnamefont{A.~C.}
  \bibnamefont{Gossard}}, \bibinfo{journal}{Phys. Rev. Lett.}
  \textbf{\bibinfo{volume}{79}}, \bibinfo{pages}{5106} (\bibinfo{year}{1997}).

\bibitem[{\citenamefont{Joo et~al.}(2006)\citenamefont{Joo, Hong, Rhie, Jung,
  Kim, Kim, Lee, Park, and Shin}}]{Joo2006}
\bibinfo{author}{\bibfnamefont{S.}~\bibnamefont{Joo}},
  \bibinfo{author}{\bibfnamefont{J.}~\bibnamefont{Hong}},
  \bibinfo{author}{\bibfnamefont{K.}~\bibnamefont{Rhie}},
  \bibinfo{author}{\bibfnamefont{K.~Y.} \bibnamefont{Jung}},
  \bibinfo{author}{\bibfnamefont{K.~H.} \bibnamefont{Kim}},
  \bibinfo{author}{\bibfnamefont{S.~U.} \bibnamefont{Kim}},
  \bibinfo{author}{\bibfnamefont{B.~C.} \bibnamefont{Lee}},
  \bibinfo{author}{\bibfnamefont{W.~H.} \bibnamefont{Park}}, \bibnamefont{and}
  \bibinfo{author}{\bibfnamefont{K.}~\bibnamefont{Shin}}, \bibinfo{journal}{J.
  Korean Phys. Soc.} \textbf{\bibinfo{volume}{48}}, \bibinfo{pages}{642}
  (\bibinfo{year}{2006}).

\bibitem[{\citenamefont{Bae et~al.}(2007)\citenamefont{Bae, Lin, Yoon, Kim,
  Bird, Imre, Porod, and Reno}}]{Bae2007}
\bibinfo{author}{\bibfnamefont{J.~U.} \bibnamefont{Bae}},
  \bibinfo{author}{\bibfnamefont{T.~Y.} \bibnamefont{Lin}},
  \bibinfo{author}{\bibfnamefont{Y.}~\bibnamefont{Yoon}},
  \bibinfo{author}{\bibfnamefont{S.~J.} \bibnamefont{Kim}},
  \bibinfo{author}{\bibfnamefont{J.~P.} \bibnamefont{Bird}},
  \bibinfo{author}{\bibfnamefont{A.}~\bibnamefont{Imre}},
  \bibinfo{author}{\bibfnamefont{W.}~\bibnamefont{Porod}}, \bibnamefont{and}
  \bibinfo{author}{\bibfnamefont{J.~L.} \bibnamefont{Reno}},
  \bibinfo{journal}{Appl. Phys. Lett.} \textbf{\bibinfo{volume}{91}},
  \bibinfo{pages}{022105} (\bibinfo{year}{2007}).

\bibitem[{\citenamefont{Lin et~al.}(2010)\citenamefont{Lin, Lim, Andrews,
  Strasser, and Bird}}]{Lin2010}
\bibinfo{author}{\bibfnamefont{T.}~\bibnamefont{Lin}},
  \bibinfo{author}{\bibfnamefont{K.}~\bibnamefont{Lim}},
  \bibinfo{author}{\bibfnamefont{A.~M.} \bibnamefont{Andrews}},
  \bibinfo{author}{\bibfnamefont{G.}~\bibnamefont{Strasser}}, \bibnamefont{and}
  \bibinfo{author}{\bibfnamefont{J.~P.} \bibnamefont{Bird}},
  \bibinfo{journal}{Appl. Phys. Lett.} \textbf{\bibinfo{volume}{97}},
  \bibinfo{pages}{063108} (\bibinfo{year}{2010}).

\bibitem[{\citenamefont{Leadbeater et~al.}(1995)\citenamefont{Leadbeater,
  Foden, Burroughes, Pepper, Burke, Wang, Grimshaw, and
  Ritchie}}]{Leadbeater1995}
\bibinfo{author}{\bibfnamefont{M.~L.} \bibnamefont{Leadbeater}},
  \bibinfo{author}{\bibfnamefont{C.~L.} \bibnamefont{Foden}},
  \bibinfo{author}{\bibfnamefont{J.~H.} \bibnamefont{Burroughes}},
  \bibinfo{author}{\bibfnamefont{M.}~\bibnamefont{Pepper}},
  \bibinfo{author}{\bibfnamefont{T.~M.} \bibnamefont{Burke}},
  \bibinfo{author}{\bibfnamefont{L.~L.} \bibnamefont{Wang}},
  \bibinfo{author}{\bibfnamefont{M.~P.} \bibnamefont{Grimshaw}},
  \bibnamefont{and} \bibinfo{author}{\bibfnamefont{D.~A.}
  \bibnamefont{Ritchie}}, \bibinfo{journal}{Phys. Rev. B}
  \textbf{\bibinfo{volume}{52}}, \bibinfo{pages}{R8629} (\bibinfo{year}{1995}).

\bibitem[{\citenamefont{Geim et~al.}(1992)\citenamefont{Geim, Bending, and
  Grigorieva}}]{Geim1992}
\bibinfo{author}{\bibfnamefont{A.~K.} \bibnamefont{Geim}},
  \bibinfo{author}{\bibfnamefont{S.~J.} \bibnamefont{Bending}},
  \bibnamefont{and} \bibinfo{author}{\bibfnamefont{I.~V.}
  \bibnamefont{Grigorieva}}, \bibinfo{journal}{Phys. Rev. Lett.}
  \textbf{\bibinfo{volume}{69}}, \bibinfo{pages}{2252} (\bibinfo{year}{1992}).

\bibitem[{\citenamefont{Carmona et~al.}(1995)\citenamefont{Carmona, Geim,
  Nogaret, Main, Foster, Henini, and Blamire}}]{Carmona1995}
\bibinfo{author}{\bibfnamefont{H.~A.} \bibnamefont{Carmona}},
  \bibinfo{author}{\bibfnamefont{A.~K.} \bibnamefont{Geim}},
  \bibinfo{author}{\bibfnamefont{A.}~\bibnamefont{Nogaret}},
  \bibinfo{author}{\bibfnamefont{P.~C.} \bibnamefont{Main}},
  \bibinfo{author}{\bibfnamefont{T.~J.} \bibnamefont{Foster}},
  \bibinfo{author}{\bibfnamefont{M.}~\bibnamefont{Henini}},
  \bibinfo{author}{\bibfnamefont{S.~P.} \bibnamefont{Beaumont}},\bibnamefont{and}
  \bibinfo{author}{\bibfnamefont{M.~G.} \bibnamefont{Blamire}},
  \bibinfo{journal}{Phys. Rev. Lett.} \textbf{\bibinfo{volume}{74}},
  \bibinfo{pages}{3009} (\bibinfo{year}{1995}).

\bibitem[{\citenamefont{Ye et~al.}(1995)\citenamefont{Ye, Weiss, Gerhardts,
  Seeger, von Klitzing, Eberl, and Nickel}}]{Ye1995}
\bibinfo{author}{\bibfnamefont{P.~D.} \bibnamefont{Ye}},
  \bibinfo{author}{\bibfnamefont{D.}~\bibnamefont{Weiss}},
  \bibinfo{author}{\bibfnamefont{R.~R.} \bibnamefont{Gerhardts}},
  \bibinfo{author}{\bibfnamefont{M.}~\bibnamefont{Seeger}},
  \bibinfo{author}{\bibfnamefont{K.}~\bibnamefont{von Klitzing}},
  \bibinfo{author}{\bibfnamefont{K.}~\bibnamefont{Eberl}}, \bibnamefont{and}
  \bibinfo{author}{\bibfnamefont{H.}~\bibnamefont{Nickel}},
  \bibinfo{journal}{Phys. Rev. Lett.} \textbf{\bibinfo{volume}{74}},
  \bibinfo{pages}{3013} (\bibinfo{year}{1995}).

\bibitem[{\citenamefont{Johnson et~al.}(1997)\citenamefont{Johnson, Bennett,
  Yang, Miller, and Shanabrook}}]{Johnson1997}
\bibinfo{author}{\bibfnamefont{M.}~\bibnamefont{Johnson}},
  \bibinfo{author}{\bibfnamefont{B.~R.} \bibnamefont{Bennett}},
  \bibinfo{author}{\bibfnamefont{M.~J.} \bibnamefont{Yang}},
  \bibinfo{author}{\bibfnamefont{M.~M.} \bibnamefont{Miller}},
  \bibnamefont{and} \bibinfo{author}{\bibfnamefont{B.~V.}
  \bibnamefont{Shanabrook}}, \bibinfo{journal}{Appl. Phys. Lett.}
  \textbf{\bibinfo{volume}{71}}, \bibinfo{pages}{974} (\bibinfo{year}{1997}).

\bibitem[{\citenamefont{Monzon et~al.}(1997)\citenamefont{Monzon, Johnson, and
  Roukes}}]{Monzon1997}
\bibinfo{author}{\bibfnamefont{F.~G.} \bibnamefont{Monzon}},
  \bibinfo{author}{\bibfnamefont{M.}~\bibnamefont{Johnson}}, \bibnamefont{and}
  \bibinfo{author}{\bibfnamefont{M.~L.} \bibnamefont{Roukes}},
  \bibinfo{journal}{Appl. Phys. Lett.} \textbf{\bibinfo{volume}{71}},
  \bibinfo{pages}{3087} (\bibinfo{year}{1997}).

\bibitem[{\citenamefont{Cerchez et~al.}(2007)\citenamefont{Cerchez, Hugger,
  Heinzel, and Schulz}}]{Cerchez2007}
\bibinfo{author}{\bibfnamefont{M.}~\bibnamefont{Cerchez}},
  \bibinfo{author}{\bibfnamefont{S.}~\bibnamefont{Hugger}},
  \bibinfo{author}{\bibfnamefont{T.}~\bibnamefont{Heinzel}}, \bibnamefont{and}
  \bibinfo{author}{\bibfnamefont{N.}~\bibnamefont{Schulz}},
  \bibinfo{journal}{Phys. Rev. B} \textbf{\bibinfo{volume}{75}},
  \bibinfo{pages}{035341} (\bibinfo{year}{2007}).

\bibitem[{\citenamefont{DeMartino et~al.}(2007)\citenamefont{DeMartino,
  Dell'Anna, and Egger}}]{Martino2007}
\bibinfo{author}{\bibfnamefont{A.}~\bibnamefont{DeMartino}},
  \bibinfo{author}{\bibfnamefont{L.}~\bibnamefont{Dell'Anna}},
  \bibnamefont{and} \bibinfo{author}{\bibfnamefont{R.}~\bibnamefont{Egger}},
  \bibinfo{journal}{Phys. Rev. Lett.} \textbf{\bibinfo{volume}{98}},
  \bibinfo{pages}{066802} (\bibinfo{year}{2007}).

\bibitem[{\citenamefont{Neto et~al.}(2009)\citenamefont{Neto, Guinea, Peres,
  Novoselov, and Geim}}]{Castro2009}
\bibinfo{author}{\bibfnamefont{A.~H.~C.} \bibnamefont{Neto}},
  \bibinfo{author}{\bibfnamefont{F.}~\bibnamefont{Guinea}},
  \bibinfo{author}{\bibfnamefont{N.~M.~R.} \bibnamefont{Peres}},
  \bibinfo{author}{\bibfnamefont{K.~S.} \bibnamefont{Novoselov}},
  \bibnamefont{and} \bibinfo{author}{\bibfnamefont{A.~K.} \bibnamefont{Geim}},
  \bibinfo{journal}{Rev. Mod. Phys.} \textbf{\bibinfo{volume}{81}},
  \bibinfo{pages}{109} (\bibinfo{year}{2009}).

\end{thebibliography}

\end{document}